\documentclass[preprint]{sig-alternate-05-2015}

\usepackage[utf8]{inputenc}
\usepackage{graphicx}
\graphicspath{{figs/}}
\usepackage[usenames,dvipsnames]{color}
\usepackage{amsmath,amssymb}
\usepackage{ifthen}
\usepackage{enumerate}
\usepackage{url}
\usepackage{xspace}
\usepackage{microtype}
\usepackage{flushend}

\begin{document}

\CopyrightYear{2016}
\setcopyright{acmlicensed}
\conferenceinfo{Middleware'16,}{December 12 - 16, 2016, Trento, Italy}
\isbn{978-1-4503-4300-8/16/12}\acmPrice{\$15.00}
\doi{http://dx.doi.org/10.1145/2988336.2988346}

\newboolean{showcomments}
\setboolean{showcomments}{true}
\ifthenelse{\boolean{showcomments}}
{ \newcommand{\mynote}[3]{
   \fbox{\bfseries\sffamily\scriptsize#1}
   {\small$\blacktriangleright$\textsf{\emph{\color{#3}{#2}}}$\blacktriangleleft$}}}
{ \newcommand{\mynote}[3]{}}

\conferenceinfo{Middleware}{'16 Trento, Italy}

\newcommand{\SYS}{SCBR\xspace}

\title{Secure Content-Based Routing Using\\Intel Software Guard Extensions}

%\subtitle{Track: Experimentation and Deployment}

\numberofauthors{4}

\author{
% 1st. author
\alignauthor Rafael Pires\\
    \affaddr{University of Neuch\^{a}tel}\\
    \email{rafael.pires@unine.ch}
% 2nd. author
\alignauthor Marcelo Pasin\\
    \affaddr{University of Neuch\^{a}tel}\\
    \email{marcelo.pasin@unine.ch}
% 3rd. author
\alignauthor Pascal Felber\\
    \affaddr{University of Neuch\^{a}tel}\\
    \email{pascal.felber@unine.ch}
\and  % use '\and' if you need 'another row' of author names
% 4th. author
\alignauthor Christof Fetzer\\
    \affaddr{TU Dresden}\\
    \email{\mbox{\!\!\!\!\!christof.fetzer@tu-dresden.de}}
}

\maketitle

%!TEX root = paper.tex
\begin{abstract}

Content-based routing (CBR) is a powerful model that supports scalable asynchronous communication among large sets of geographically distributed nodes.
Yet, preserving privacy represents a major limitation for the wide adoption of CBR, notably when the routers are located in public clouds.
Indeed, a CBR router must see the content of the messages sent by data producers, as well as the filters (or subscriptions) registered by data consumers.
This represents a major deterrent for companies for which data is a key asset, as for instance in the case of financial markets or to conduct sensitive business-to-business transactions.
While there exists some techniques for privacy-preserving computation, they are either prohibitively slow or too limited to be usable in real systems.
In this paper, we follow a different strategy by taking advantage of trusted hardware extensions that have just been introduced in off-the-shelf processors and provide a trusted execution environment.
We exploit Intel's new \emph{software guard extensions} (SGX) to implement a CBR engine in a \emph{secure enclave}.
Thanks to the hardware-based trusted execution environment (TEE), the compute-intensive CBR operations can operate on decrypted data shielded by the enclave and leverage efficient matching algorithms.
Extensive experimental evaluation shows that SGX adds only limited overhead to insecure plaintext matching outside secure enclaves while providing much better performance and more powerful filtering capabilities than alternative software-only solutions.
To the best of our knowledge, this work is the first to demonstrate the practical benefits of SGX for privacy-preserving CBR.
\end{abstract}


\newpage
%!TEX root = paper.tex
\section{Introduction}
\label{sec:introduction}

Content-based routing (CBR) is a flexible and powerful paradigm for scalable communication among distributed processes.
It decouples data producers from consumers, and dynamically routes messages based on their content.
While the publish/subscribe communication model has been extensively studied over more than a decade \cite{PS:Survey:03}, and many implementations have been proposed (e.g., \cite{Jacobsen:USENIX:2007, Choi2010, Nabeel:SACMAT:2012}), it still fails to reach wide deployment and usage in the era of cloud computing.

One of the major reasons to the lack of general adoption can be tracked down to privacy concerns.
Indeed, CBR requires router components to filter messages by matching their content against a (potentially large) collection of subscriptions that act as a reverse index and, hence, must be stored by the filtering engines.
In turn, this requires the router to see the content of both the messages and the subscriptions, which represents a major threat for companies for which data is a key asset.
For instance, in the emblematic example of financial trading, stock quotes published by exchange platforms have commercial value and must be protected, while subscriptions may reveal sensitive information about a client's portfolio and must also be secured.

Several approaches to supporting privacy-preserving publish/subscribe have been proposed recently \cite{PPPS:Survey:16, Uzunov:CompSec:2016}, but none provides at the same time powerful filtering capabilities and high performance.
Approaches that rely on sophisticated cryptographic techniques like (fully) homomorphic encryption can support a wide range of operations but are prohibitively slow---several orders of magnitude slower than plaintext filtering.
Performance can be improved by exploiting specialised techniques like asymmetric scalar-product preserving encryption (ASPE) \cite{Choi2010}, but these approaches suffer from severe limitations in the type of operations they support, which is typically restricted to equality matching or degraded forms of range queries.
Furthermore, these techniques also often introduce additional overheads, e.g., ASPE's space complexity grows exponentially with the number of attributes.

In order to combine both security and performance, one can instead resort on hardware-based secure execution environments.
However, such extensions have been so far limited to narrow, domain-specific applications (e.g., military) and unsuitable for deployment in cloud environments, notably because of their high cost and dependency on custom hardware.
This is about to change thanks to availability of trusted execution environments (TEEs) in commodity processors.
In particular, processors built using Intel's new Skylake microarchitecture support a new extension set, called \emph{software guard extensions} (SGX), that allows users to execute cryptographically-signed applications in a \emph{secure enclave} that is shielded from other code running on the same processor.
Data is encrypted while outside the enclave, but can be processed in plaintext form (and hence with almost no performance overhead) inside the enclave.

The first SGX-enabled processors have shipped in late 2015 in a mobile version, and desktop and server variants have followed in 2016.
Hence this technology is very new and still largely untested in real-world settings.
In this paper, we introduce an original CBR architecture that exploits the SGX technology to execute a routing engine in a secure enclave.
We propose a protocol for securely exchanging cryptographic keys between data producers and consumers, and the SGX-protected routing engine.
Both publications and subscriptions are encrypted and signed, thus protecting the system from unauthorised parties observing or tampering with the information.
Our system, called \SYS, thus combines a key exchange protocol and a state-of-the-art routing engine to provide both security and performance while executing under the protection of the secure enclave.
To the best of our knowledge, \SYS is the first system to experimentally evaluate and demonstrate the benefits of executing CBR in a TEE.

In this paper, we propose the following contributions.
We first present a detailed description of the SGX technology and how it can be used to develop secure implementation ($\S$\ref{sec:background}).
We then introduce an original secure CBR architecture that combines operations on encrypted data outside enclaves, and efficient matching of publications and subscriptions inside enclaves ($\S$\ref{sec:architecture}).
We conduct in-depth evaluation on a real implementation with several workloads to observe the sources of performance overheads and the various trade-offs of SGX, and we provide comparative results against plaintext (insecure) matching as well as an ASPE-based alternative ($\S$\ref{sec:evaluation}).
The paper is completed by a discussion of related work ($\S$\ref{sec:related}) and concluding remarks ($\S$\ref{sec:conclusion}).

%!TEX root = paper.tex
\section{Background}
\label{sec:background}

\subsection{Intel Software Guard Extensions (SGX)}

Traditionally, one protects the integrity and confidentiality of applications by enforcing the isolation of applications.
An operating system isolates the applications using hardware mechanisms like virtual address spaces and privileged instructions.
Multiple operating systems running on the same physical host are isolated by the hypervisor using hardware virtualisation extensions provided by the CPU.  

The protection of the application integrity and confidentiality requires so far that both the operating system as well as the hypervisor are trusted.
In cloud environments, at least the hypervisor and, in some cases, also the operating system are under the control of the cloud provider. 
Trusting the hypervisor and the operating system in cloud environments can raise legal and technical issues since the cloud provider is a different legal entity from the application provider.

\begin{figure}
\centering
\includegraphics[scale=0.8]{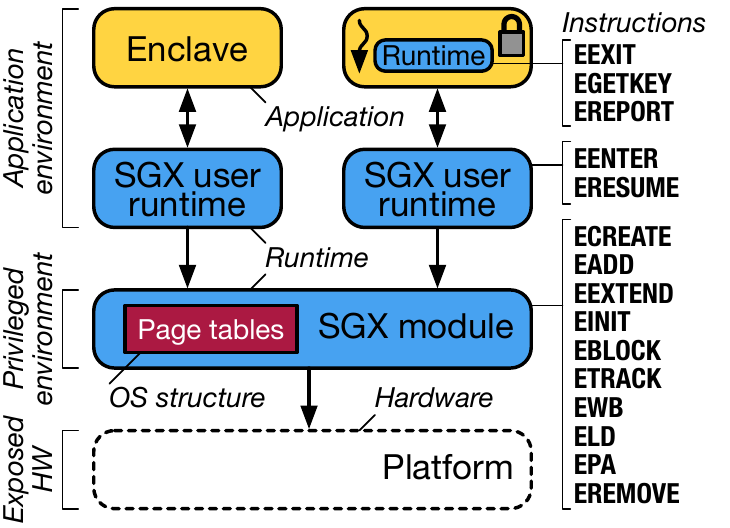}
\caption{\label{fig:sgx-arch}
SGX enclaves are a trusted execution environment for unprivileged applications.
Enclaves can only be created with the help of the operating system.
In this way, the operating system can control which applications are permitted to create enclaves as well as the number and the size of the enclaves.}
\end{figure}

From a legal point of view, contractual agreements with the cloud provider could ensure legal protection of the data integrity and confidentiality.
However, such agreements might be insufficient in case the cloud provider is located in a different jurisdiction. 
Moreover, from a technical perspective, operating systems and hypervisor encompass millions of lines of source code.
The number of exploitable bugs is proportional to the number of lines of code.
Hence, on shared infrastructures, one cannot neglect the probability of applications being attacked by other tenants' applications executing on the same computer. 
With formal proofs of the operating system \cite{Klein:2009:SFV:1629575.1629596}, one might be able to reduce the number of exploitable bugs.
Yet, the system administrators of the cloud provider still have access to all application data and, therefore, their credentials are often hijacked to gain access to systems and all application data being processed \cite{sysadmins}. 

One needs a technology that protects the confidentiality and integrity of application data from access by any other software, even software with higher privileges like the hypervisor.
Trusted execution environments (TEE) are a way to protect the integrity and confidentiality of data.
ARM TrustZone \cite{arm2009trustzone} is a popular trusted execution environment that provides a \emph{secure world} that cannot be accessed by the \emph{normal world}.
It provides, however, only one secure world. 
Hence, applications either need to share the secure world, or at most one application can use the secure world.
Moreover, the secure world is under the control of a separate operating system, which means that it can still be accessed by some system administrator.

Intel SGX is a recently released technology that provides TEEs implemented by the CPU (see Figure~\ref{fig:sgx-arch}).
Although TEE solutions have been proposed and implemented previously, SGX has the advantage that each application can create separate \emph{secure enclaves} to protect the confidentiality and integrity of its data while it is being processed.
This closes the gap of traditional approaches that protect the data using protocols like transport layer security (TLS) \cite{rfc5246} during transmission, and using file encryption when storing it---but so far do not protect it while it is processed.

The protection of data during processing is particularly important in the context of cloud computing, where applications run in an environment under the control of a different legal entity.
Developers can ensure the end-to-end confidentiality and integrity of their application's data by terminating the TLS connections that connect the application with its clients inside of an enclave, and any data at rest remains encrypted inside the enclave.
With this approach, neither the cloud provider nor any hacker with root access can compromise the integrity or confidentiality of application data.
 
\begin{figure}
\centering
\includegraphics[scale=0.8]{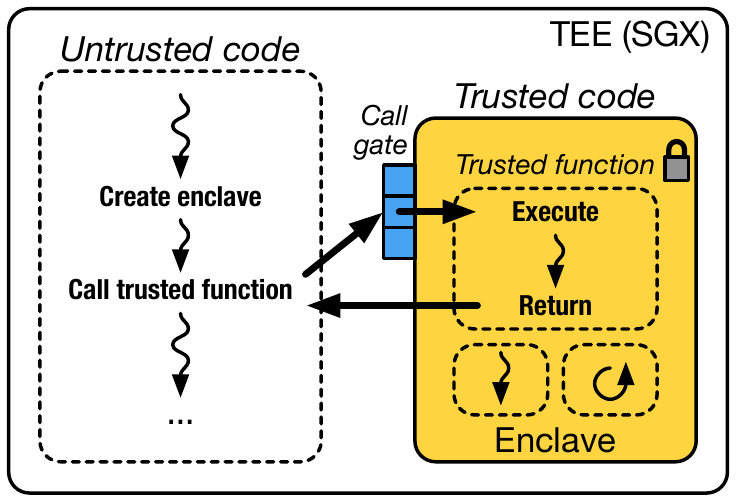}
\caption{\label{fig:sgx-call}
Sensitive data can be protected inside of an enclave.
This data can be accessed by calling trusted functions inside of the enclave.}
\end{figure}

The basic idea underlying SGX is that an application can keep its confidential data inside of an enclave and access it from within (see Figure \ref{fig:sgx-call}).
For example, an enclave could protect the credentials of an external database and additionally ensure that only certain queries are issued, or limit the maximum frequency of the database queries sent by this application.

For small applications, SGX can protect application data even when the attacker has physical access to the computer.
The hardware security perimeter is the CPU package, and all data belonging to an enclave is encrypted and authenticated when stored in main memory.
External snooping, such as eavesdropping the memory bus or the system memory itself, will hence not reveal any data stored inside of enclaves.
All data read from memory is checked for integrity and freshness by verifying the authentication tags. 
External data modifications are therefore detected since the authentication codes will not match. 
Feeding the CPU with old data that was properly encrypted and authenticated is also detected by keeping track of the latest authentication codes for each page.

To enable an application to use enclaves, the developer must provide a signed shared library (\texttt{.so} or \texttt{.dll}) that will execute inside an enclave. 
The library itself is not encrypted and can be inspected before being started, hence no secret should be stored inside the code. 

An enclave is provided with secrets, like certificates and keys, with the help of a remote attestation protocol.
This protocol can prove that an enclave runs on a genuine Intel's processor with SGX and verify that its identity matches that of the code that the developer asked to start \cite{SgxMemory2016}.
During remote attestation, a secure channel is established that permits the remote entity to provide the enclave with secrets.

Enclave code and data are stored in a memory area predefined at boot time, called the enclave page cache (EPC), which is at the moment limited to 128\;MB.
Applications can use approximately 90\;MB while the remaining space is reserved for SGX itself.
If an enclave is larger than 90\;MB, any access to an enclave page that does not reside in the EPC results in a page fault.
The page fault is handled by an SGX driver in the operating system that selects a page of the EPC to \emph{evict}, i.e., the page is moved to main memory in case it was \emph{dirty} or just dropped if a copy already exists in main memory.
After a page is evicted, the SGX driver loads the requested page (that triggered the fault) from main memory.
The SGX driver closely interacts with the CPU for eviction and paging since the CPU keeps track of the authentication tags of the evicted pages and checks them when loading pages into the EPC.
In this way, one does not need to trust the SGX driver since it cannot violate the confidentiality nor the integrity of enclave pages.

Confidentiality in the traffic between CPU and system memory is achieved by a component called the memory encryption engine (MEE), which is also responsible for providing tamper resistance and replay protection. 
Under normal processor operation, memory transactions that miss the cache are handled by the memory controller (MC).
If however the cache miss is translated to a protected region, MEE takes over. 
In this case, it encrypts or decrypts data before sending to or fetching from system memory, in addition to performing integrity checks \cite{SgxMemory2016}.

Memory checks are made through an integrity tree \cite{IntegrityTreeSurvey2009} that uses a stateful message authentication code (MAC) with nonces \cite{StatefulIntegrityTrees2007} (non repeating numbers, coined to be used only once).
The tree is stored in untrusted memory, except for the root that is kept on-die and inaccessible from outside.
It reflects the integrity of the whole protected area at a given time, thus precluding any attacks by means of modification or replay of values in memory.
Any mismatch during a verification causes a MC lock, which ultimately requires a machine reboot \cite{SgxMemory2016}.

The SGX system provides each enclave with a \emph{seal key} that can be used to store data on stable storage and access it again upon subsequent execution.
This facilitates the development of applications that can restart an enclave without requiring a new remote attestation.
The enclave instead loads its secrets from a configuration file encrypted with the enclave-specific seal key and kept in stable storage.
Note that an attacker could still try to serve an enclave with a previous version of a configuration file that is properly encrypted and authenticated.
To prevent such replay attacks, an enclave can use the monotonic counter facilities provided by the platform.
Each time an enclave writes a new version of its configuration data, it increments a monotonic counter and stores the new value inside the configuration file.
When the enclave restarts, it reads the monotonic counter and checks that it matches the values stored in the configuration file.

%!TEX root = paper.tex
\section{\SYS Architecture}
\label{sec:architecture}

We describe in this section the overall architecture of \SYS.
We first explain the rationale that drove the design of the system.
We then introduce the considered model, and we present the principle of subscription registration and message publication.
We finally discuss some implementation details.

\subsection{Objectives and Assumptions}

Designing a secure, privacy-preserving CBR system is not trivial, even if one can rely on trusted hardware.
Consider the example of a stock exchange publishing streams of financial data (stock quotes).
Consumers who want to protect their confidential data (portfolio) have to trust the code of the CBR engine, which typically originates from the stock exchange.
They may also want to restrict the ability to see their subscriptions to a single publisher, and not other data providers that leverage the same software and infrastructure (e.g., different stock exchange operators).

We therefore consider a simple model based on the following assumptions.
The publish/subscribe system operates as a service under the control of a single ``service provider'' that publishes data.
Consumers are the clients of the service and typically pay recurring fees to have access to the data.
Hence, there must be ways for the producers to control which subscribers can join and read data from the service, and to exclude clients that stop paying their fees or behave in a non-trustworthy manner.
The publishers operate within the administrative domain of the service provider from which data originates, and they are trusted by the clients for the purpose of the considered service.
This model closely maps, for instance, to the aforementioned scenario of the stock exchange.

Given the trust relationships between the different components of the system, it appears clear that publishers and clients must share cryptographic keys that are not known by the infrastructure.
Furthermore, there should be some lightweight mechanisms for the publishers to (dis-)allow clients from accessing new data, independently of whether they had been legitimate customers in the past.

\begin{figure}
\centering
\includegraphics[scale=0.8]{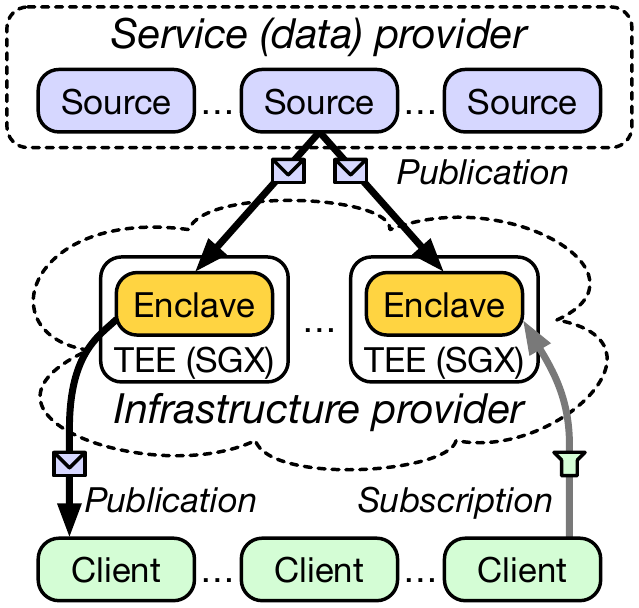}
\caption{\label{fig:roles}Components and roles.}
\end{figure}

\subsection{System Model}

Following the considerations discussed above, we distinguish three main roles in our architecture, as illustrated in Figure~\ref{fig:roles}:
\begin{itemize}
\item
The \emph{service provider} (or \emph{data provider}) produces information flows for the clients, typically ``as a service'' and for a fee.
The data may be produced by multiple sources (publishers) operating within the same administrative domain.
\item
The \emph{infrastructure provider} hosts the CBR engines in the cloud.
It provides secure hardware (SGX-based enclaves) and performs the actual data routing and transmission through its network.
As it operates under a different administrative domain and may share its resources among several customers (in a multi-tenant configuration), the infrastructure provider is trusted neither by the data provider nor by its clients.
\item The \emph{clients} of the service are the end users who are interested in the actual data and subscribe to information flows via the CBR engines.
They trust the data providers but not the infrastructure.
\end{itemize}

Messages are composed of a payload, which is of interest to the end users but opaque to the \SYS system, and a header that contains several \emph{attributes} and associated values.
The CBR engines filter messages based on the attribute values in their header.

Subscriptions are composed of \emph{predicates} specifying \emph{constraints} over the attributes.
Predicates can include equality constraints or generally any kind of ranges over the values of the attributes.
For instance, a subscriber interested in specific quotes for a company when it reaches a certain price can register a subscription such as ``$symbol = \texttt{"HAL"} \wedge price < 50$''.
We say that a message \emph{matches} a subscription if its header satisfies the constraints expressed in the subscription predicate.

Subscriptions are typically stored by the CBR engine in a dedicated data structure that operates as an ``inverted'' database.
By exploiting relationships between the different predicates, as pioneered in \textsc{Siena} \cite{Carzaniga2001}, one can both reduce the memory footprint of the subscription index and improve the matching speed.
In particular, the property of \emph{containment} (also called covering) can be leveraged to avoid unnecessary tests.
Essentially, we say that a subscription $s$ contains or covers another subscription $s'$, denoted by $s \sqsupseteq s'$, if any event that matches $s'$ also matches $s$.
That is, $s$ is more general than $s'$.
For instance, predicate ``$x>0$'' covers both predicates ``$x=1$'' and subscription ``$x>0 \wedge y=1$''.
Note that the containment relationship creates a partial order on subscriptions.
In \SYS, we use a matching algorithm that exploits containment to minimise the footprint of the subscriptions stored in the enclave, where only a limited amount of memory is available.

\SYS makes use of both symmetric and asymmetric (public key) cryptography.
The former is more efficient and is used for communication between the publishers and the enclaves, while the latter is used between clients and the service provider when registering subscriptions, as will be detailed later.

\begin{figure}
\centering
\includegraphics[scale=0.8]{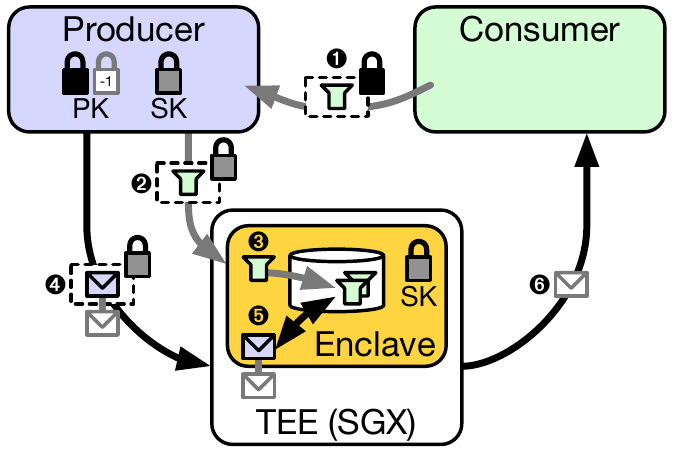}
\caption{\label{fig:interactions}Interaction amongst entities.}
\end{figure}

\subsection{The Subscription Process}

As mentioned above, \SYS was designed so that producers are the owners of the generated data.
They have therefore the ability to decide whether they accept a subscription from a client, as well as to subsequently invalidate it.
To control access to the service, we rely upon an additional admission phase when registering a new subscription.
The client cannot freely submit its subscriptions to the CBR engines in the cloud, but has to go through a data producer.
Informally, the registration process works as follows (see Figure~\ref{fig:interactions}).

Consider a client $c$ that wants to register a subscription $s$ by the routing engine $r$ and subsequently receive message $m$ sent by the data producer $p$.
The publisher has a public/private key pair ($\mathit{PK}$/$\mathit{PK}^{-1}$), as well as symmetric key ($\mathit{SK}$) that is shared with the code running in the enclave, but unknown to clients and to the infrastructure provider---this is made possible thanks to SGX, as explained in $\S$\ref{sec:background}.
\begin{enumerate}
\item
The client first encrypts its subscription $s$ using the data provider's public key, hence preventing unauthorised parties to see it, and sends the resulting encrypted subscription $\{s\}_{\mathit{PK}}$ to $p$.
\item
Then, after decrypting and verifying that the subscription is valid, as well as verifying the client's status, the publisher re-encrypts $s$ using $\mathit{SK}$ and signs it.
It then sends the encrypted subscription $\{s\}_{\mathit{SK}}$ to the routing engine $r$.
\item
Finally, $r$ validates and decrypts the subscription inside the enclave (remember that $\mathit{SK}$ is only known to the code running within the enclave), and inserts it in its index.
\end{enumerate}

Note that the subscriptions also embed information about the clients that it visible to the code running outside the enclave.
This allows the router to establish connexions with the consumers and forward relevant messages to them.

\subsection{The Publication Process}

Once subscriptions have been registered by the clients, data can be routed along the reverse path.
The publication process works as follows.
\begin{enumerate}
\setcounter{enumi}{3}
\item
The publisher encrypts the header of the message $m$ using $\mathit{SK}$, which is only known to the code running inside the enclave (encryption of the payload is discussed below).
The encrypted message $\{m\}_{\mathit{SK}}$ is then sent to the routing engine $r$.
\item
Upon receiving the message, $r$ decrypts the header in the enclave, leaving the opaque payload outside, and matches it against its subscription index.
The result of this operation is a list of clients that have registered a matching subscription.
\item
Finally, $r$ forwards the encrypted message payload to all clients that have been identified as part of the matching operation.
\end{enumerate}

The payload of messages is encrypted separately using a (symmetric) group key shared between the publisher and the consumers.
This allows publishers to periodically change the key as the population of customers evolves.
In particular, this enables publishers to prevent clients that have cancelled their membership from accessing newly published messages.
This process is orthogonal to the encryption performed for protecting the header and subscriptions, and used for privacy-preserving CBR.
It is hence out of the scope of this paper and not discussed further.

Note that having multiple routers in the path would increase the complexity of the key management between publishers and matchers. 
We believe that an overlay broker network is not the best architecture for a scalable privacy-preserving pub/sub engine, and we would rather advocate a similar structure as in StreamHub \cite{StreamHub2013} where we specialise the system components in order to gain performance. 
In such an architecture, the current publisher-matcher key management scheme could be simply replicated.

\subsection{Implementation Details}

Intel provides a set of tools to aid the coding and the fulfilment of SGX requirements. 
Their software development kit (SDK) comprises a generator for proxies and stubs written in C that should be linked to both trusted and untrusted code.  
This generation is based on a text-based configuration file that follows the syntax of the \emph{enclave definition language}, which basically defines the interface of the edge routines.  
Since system calls and input/output instructions are not allowed inside secure containers, Intel also provides libraries that are guaranteed to comply with these limitations.
We used Intel's STL implementation and AES-CTR encryption functions for enclave code.

The SDK also includes the \emph{enclave signing tool} responsible for the measurement and signature ($\S$\ref{sec:background}) of the shared library that will be loaded in the protected container \cite{SgxProgramming2014}.  
Intel's SDK is provided both for Windows and Linux, and in this work, we used the latter version.

In \SYS, encryption outside the enclave is implemented using the Crypto++ library,\footnote{\url{http://www.cryptopp.com/}} using respectively AES-CTR and RSA for symmetric and asymmetric encryption.
We use the ZeroMQ library\footnote{\url{http://zeromq.org/}} for communication, and we serialise both plain-text or encrypted messages in Base64 text format.
Information about page faults is obtained via the Linux system's \texttt{getrusage} function (attribute \texttt{minflt}).
Similarly, we rely on a Linux system call to configure and read the processor's performance counters for cache misses.

%!TEX root = paper.tex
\section{Evaluation}
\label{sec:evaluation}

We evaluated the system as described in Figure~\ref{fig:interactions}, with both the producer and consumer running in one machine and the filtering engine in another.
Measurements were collected at the machine running the filter, which was equipped with an Intel Skylake CPU model i7-6700 running at 3.4\;GHz with an 8\;MB cache and 8\;GB of main memory.
We allocated 128\;MB of main memory to EPC (maximum allowed).

\newcommand{\rbx}[1]{\raisebox{1.2ex}[0pt]{#1}}
\begin{table}[t!]
\scriptsize
\setlength\tabcolsep{2pt}
\renewcommand{\arraystretch}{1.1}
\begin{center}
\begin{tabular}{r|rcl|l|l}
\multicolumn{1}{c|}{\textbf{Workload}} & \multicolumn{3}{c|}{\textbf{Proportion of}}  & \multicolumn{1}{c|}{\textbf{Number of}}  & \multicolumn{1}{c}{\textbf{Subs values}} \\[-1pt]
\multicolumn{1}{c|}{\textbf{name}}     & \multicolumn{3}{c|}{\textbf{equality predicates}}          & \multicolumn{1}{c|}{\textbf{attributes}} & \multicolumn{1}{c}{\textbf{distribution}} \\
\hline
\texttt{e100a1}        & 100\% & : & 1 eq. pred.                  &                       &                \\
\cline{1-1}            \cline{2-4}
\texttt{e80a1}         & \multicolumn{3}{l|}{}                    & \rbx{8--11 (original)} &                \\
\cline{1-1}                                                       \cline{5-5}
\texttt{e80a2}      & \rbx{20\%} & \rbx{:} & \rbx{0 eq. pred.} & $2\times$ more        &                \\
\cline{1-1}                                                       \cline{5-5}
\texttt{e80a4}      & \rbx{80\%} & \rbx{:} & \rbx{1 eq. pred.} & $4\times$ more        &                \\
\cline{1-1}            \cline{2-4}                                \cline{5-5}
                     & 15\% & : & 0 eq. pred.                   &                       & Uniform        \\
\rbx{\texttt{extsub2}} & 60\% & : & 1 eq. pred.                   & \rbx{$2\times$ more}  &                \\
\cline{1-1}                                                       \cline{5-5}
                     & 15\% & : & 2 eq. pred.                   &                       &                \\
\rbx{\texttt{extsub4}} & 10\% & : & 3 eq. pred.                   & \rbx{$4\times$ more}  &                \\
\cline{1-1}            \cline{2-4}                                \cline{5-5}           \cline{6-6}
\texttt{e80a1z100}       & 20\% & : & 0 eq. pred.                   &                       & Zipf on symbol \\
\cline{1-1}                                                                             \cline{6-6}
\texttt{e80a1zz100}      & 80\% & : & 1 eq. pred.                   & 8--11 (original)       & Zipf on all    \\
\cline{1-1}            \cline{2-4}
\texttt{e100a1zz100}     & 100\% & : & 1 eq. pred.                  &                       & attributes     \\
\end{tabular}
\vspace{-3pt}
\caption{Workloads description (adapted from \cite{DEBS-PF:12}).}
\label{tab:workloads}
\end{center}
\end{table}

%Describe Workloads
To evaluate \SYS and to facilitate comparison, we reused the workloads from previous work \cite{DEBS-PF:12} composed of 9 datasets.
They were built based on real data corresponding to randomly selected stock quotes from the Yahoo!\ finance website.\footnote{\url{http://finance.yahoo.com/}}
Approximately 250,000 entries were collected in a period of 5 years, with publications composed by 8 to 11 attributes.
The entries collected were used to produce synthetic subscription datasets containing an assortment of equality and range predicates on the quotes' attributes.
Besides, subscriptions were selected according to (1)~a uniform random distribution and (2)~a Zipfian law with exponent $s=1$.
In order to assess the algorithms' performance with a greater number of variables and different levels of containment, some workloads were synthesised with twice and four times the number of attributes of the original publications, by merging data from multiple quotes.
Table~\ref{tab:workloads} summarises the characteristics of the datasets used.

\begin{figure}
\centering
\includegraphics[scale=0.6]{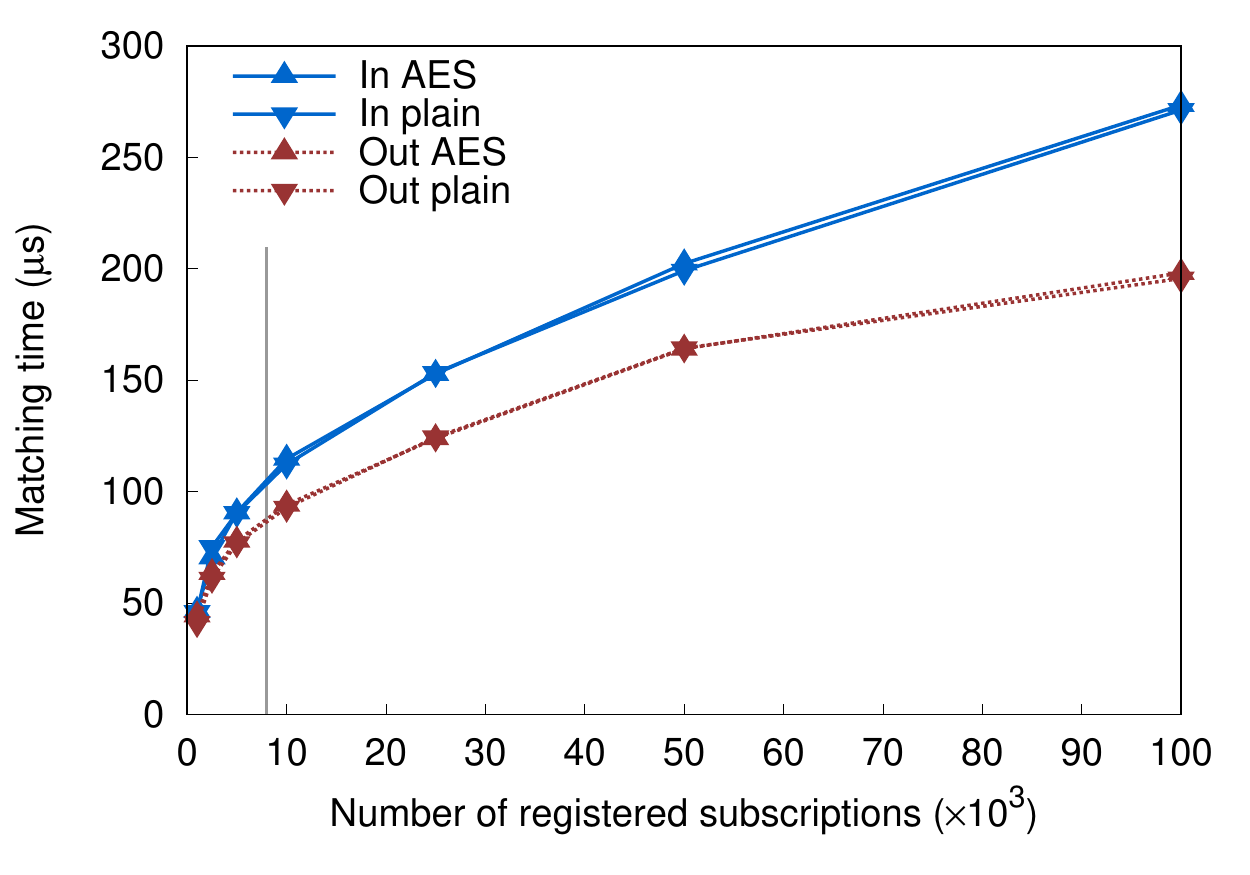}
\caption{\label{fig:encoverhead}Overhead of encryption and enclave.}
\end{figure}

Our first experiment aimed at evaluating the performance overhead caused by executing our filter inside an enclave.
We filled the subscription database with datasets with 1,000, 2,500, 5,000, 10,000, 25,000, 50,000 and 100,000 subscriptions, reaching a total memory size of approximately 43MB for the largest dataset.
Thereafter, we sent a batch of 1,000 publications to be matched against the subscriptions and measured the time it took to accomplish each filtering operation.
We ran an identical setup with and without encryption, inside and outside an enclave, using the same filtering code.
When using encryption, publications and subscriptions were encrypted in the producer and decrypted in the filter using AES-CTR.
The average results for the first workload (\emph{e100a1}) are shown in Figure~\ref{fig:encoverhead}.
By considering the proximity of the lines with and without encryption, we can see that encryption overhead is small and nearly constant.
Indeed, this overhead remains below 5$\mu s$ both inside or outside the enclave, which is negligible when compared to the matching time given a reasonable large database size.
The overhead resulting from the enclave is more significant, reaching nearly 40\% for the largest set of subscriptions considered in our experiments,
which is explained by the occurence of cache misses (cache size is shown by the vertical line).

\begin{figure}
\centering
\includegraphics[scale=0.6]{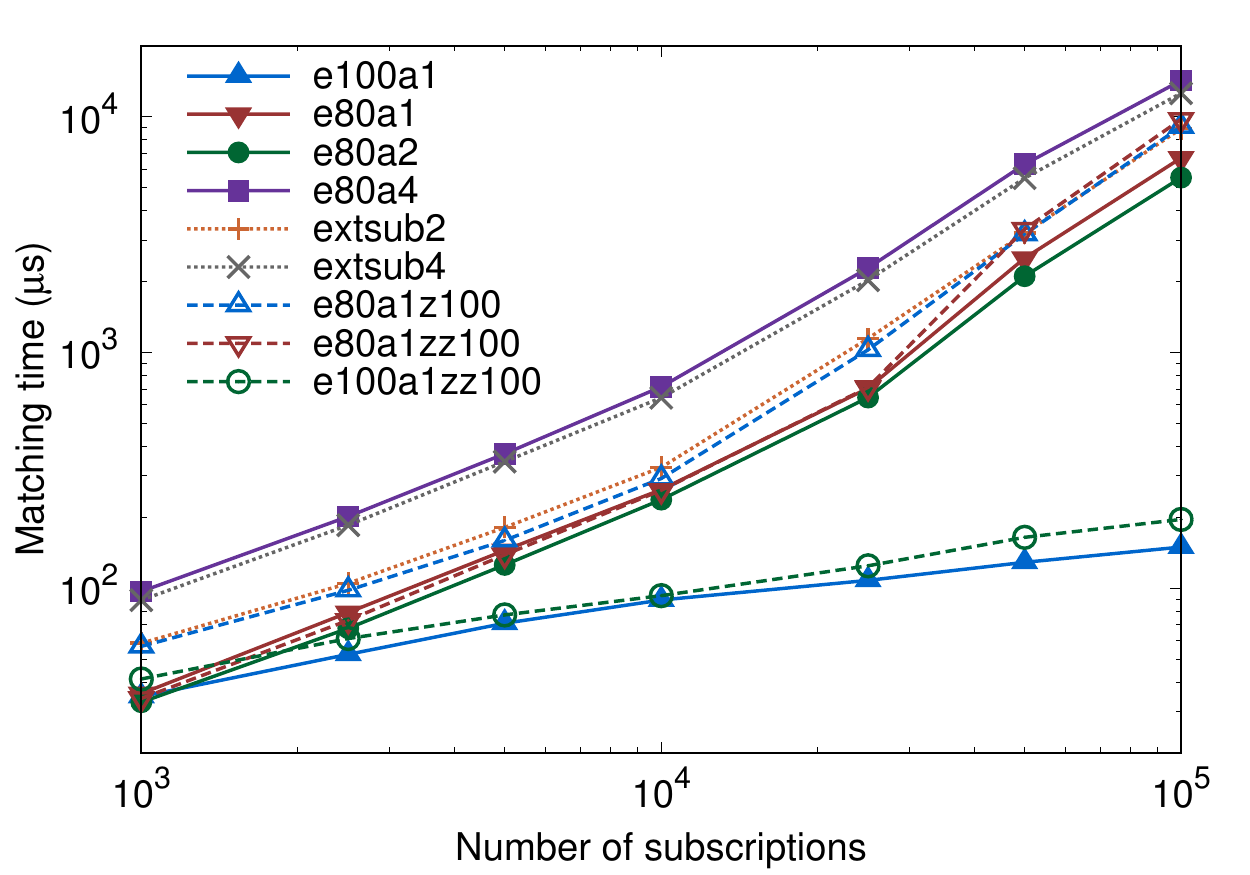}
\caption{\label{fig:workloads}Performance of the containment-based algorithm applied to the different workloads in plaintext, outside enclaves.}
\end{figure}

We then focused on the influence of the workloads.
In order to understand the effect of different datasets on \SYS performance, we first executed each of them without encryption outside secure enclaves.
Results are shown in Figure~\ref{fig:workloads}. 
The first (\emph{e100a1}) and last (\emph{e100a1zz100}) workloads show the best performance, since all subscriptions contain equality predicates and the subscription set forms deeper containment trees. 
In contrast, datasets with more attributes (\emph{e80a4} and \emph{extsub4}) perform worse because they yield indexes with more roots and shallow trees, therefore inducing more comparisons to traverse the whole database.

\begin{figure*}[t!]
\setlength{\tabcolsep}{0pt}
\centering
\begin{tabular}{ccc}
\multicolumn{3}{c}{\includegraphics[scale=0.6]{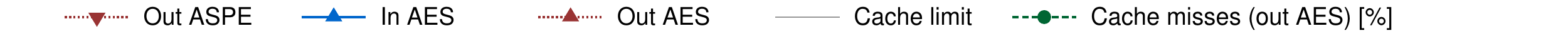}}
\\[-1mm]
\includegraphics[scale=0.6]{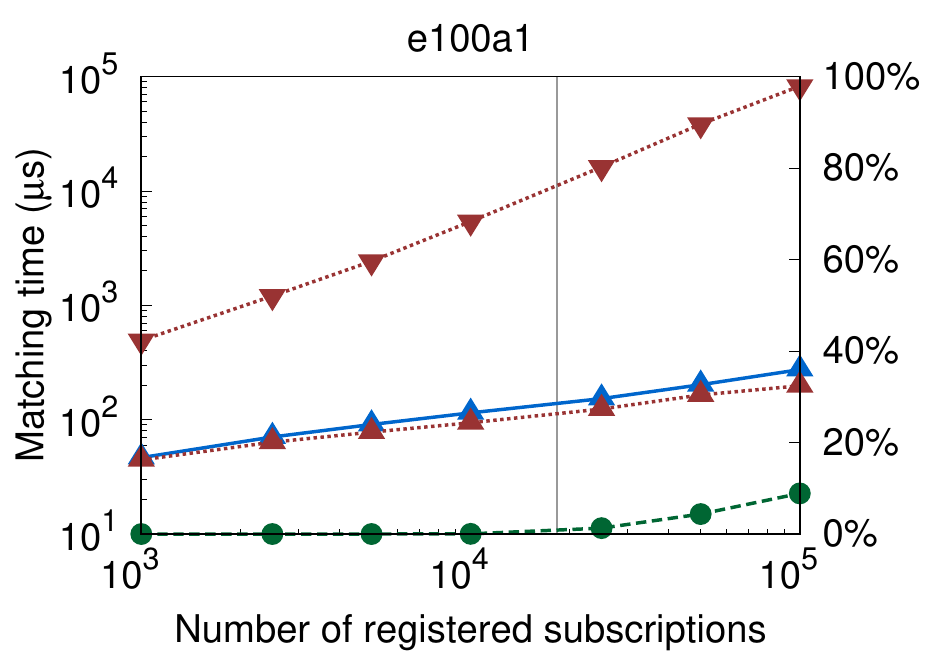}
&
\includegraphics[scale=0.6]{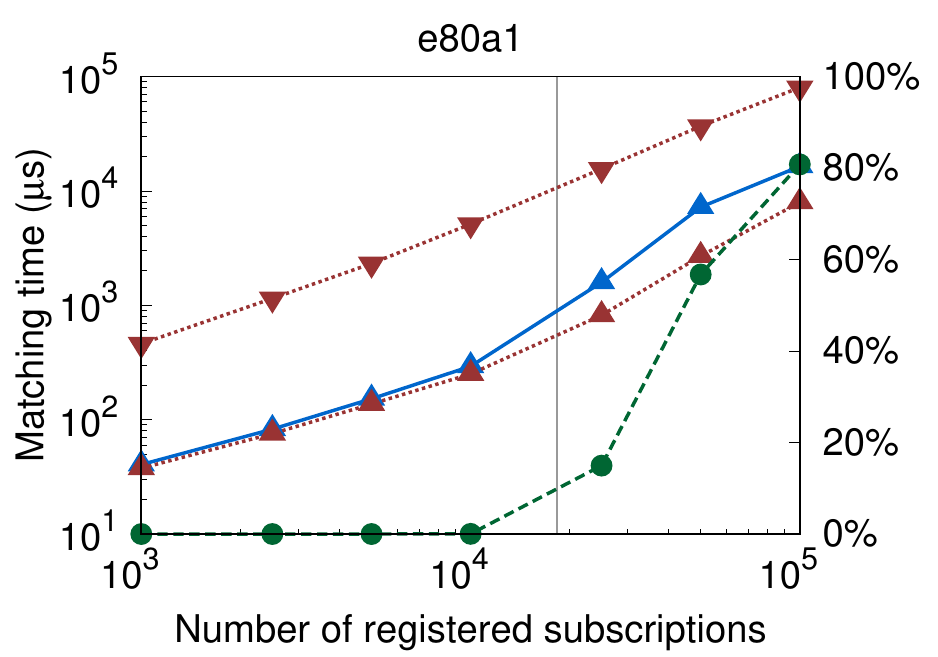}
&
\includegraphics[scale=0.6]{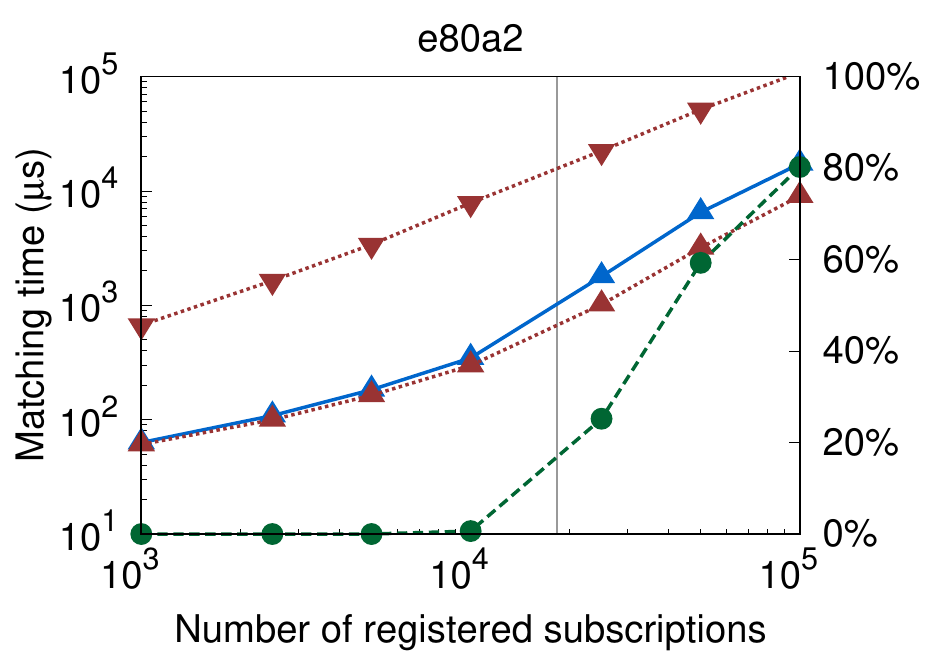}
\\[2mm]
\includegraphics[scale=0.6]{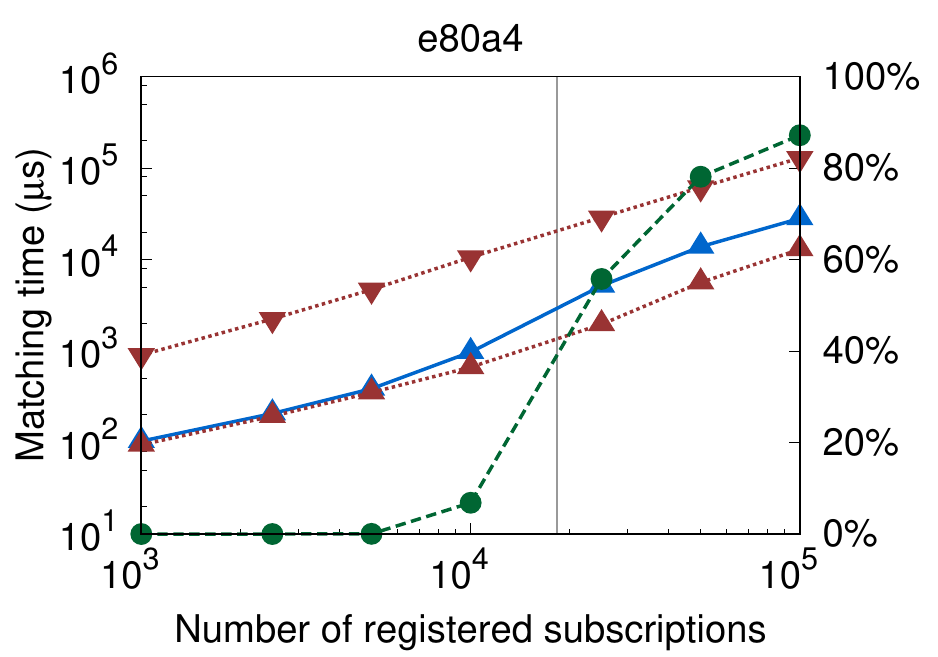}
&
\includegraphics[scale=0.6]{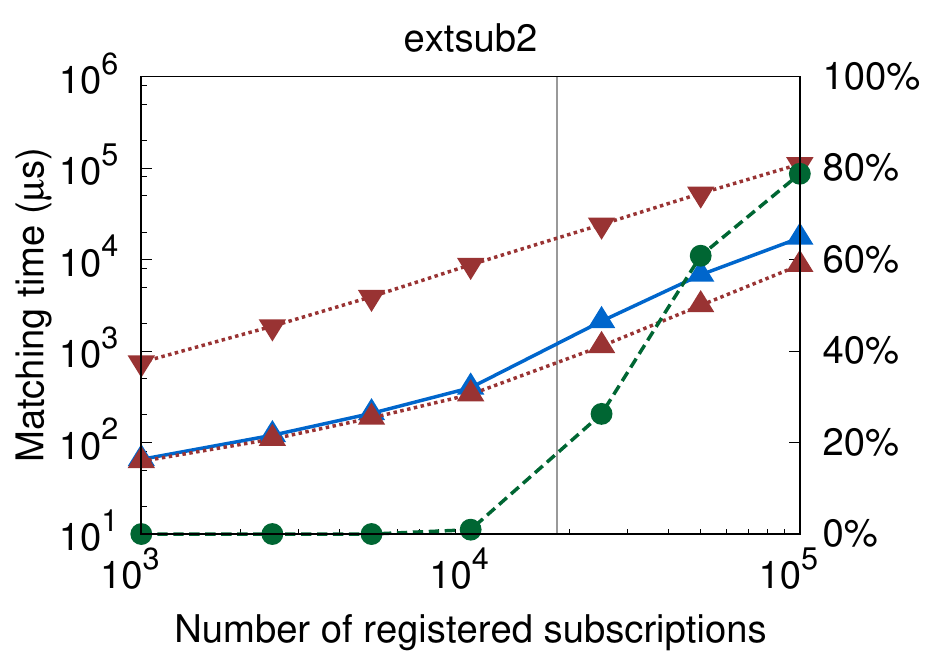}
&
\includegraphics[scale=0.6]{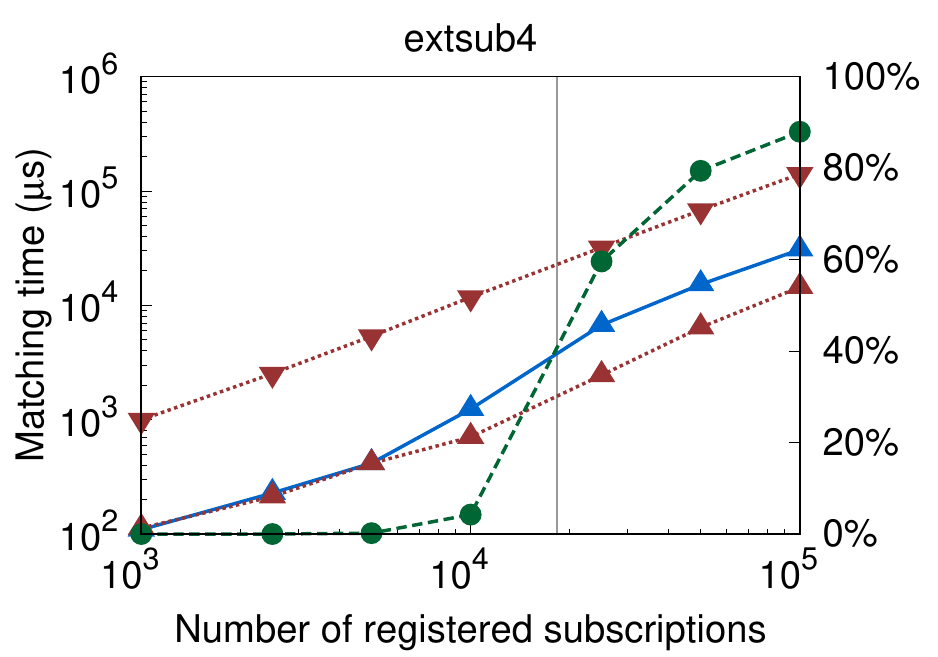}
\\[2mm]
\includegraphics[scale=0.6]{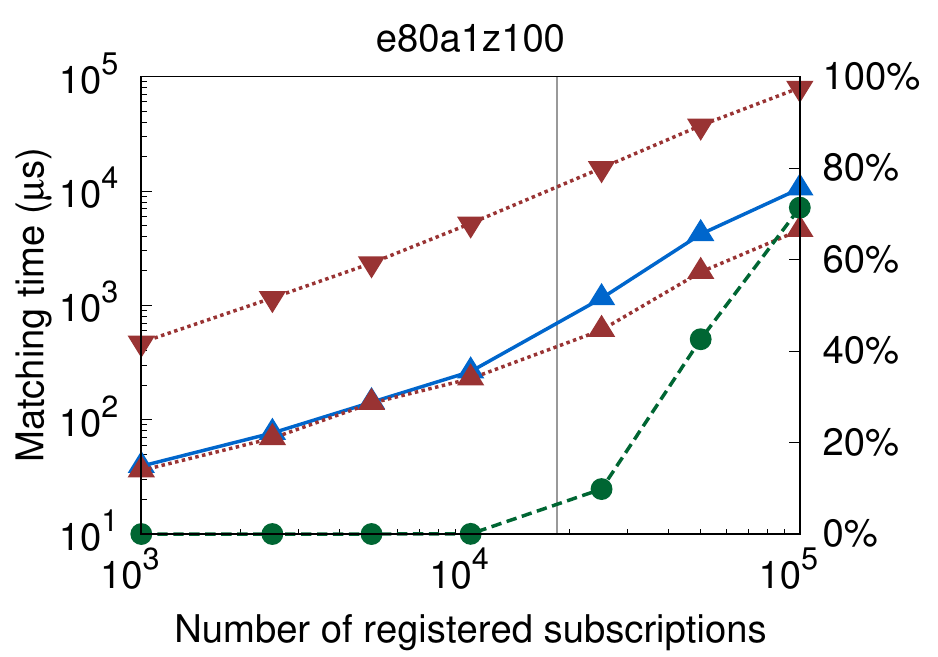}
&
\includegraphics[scale=0.6]{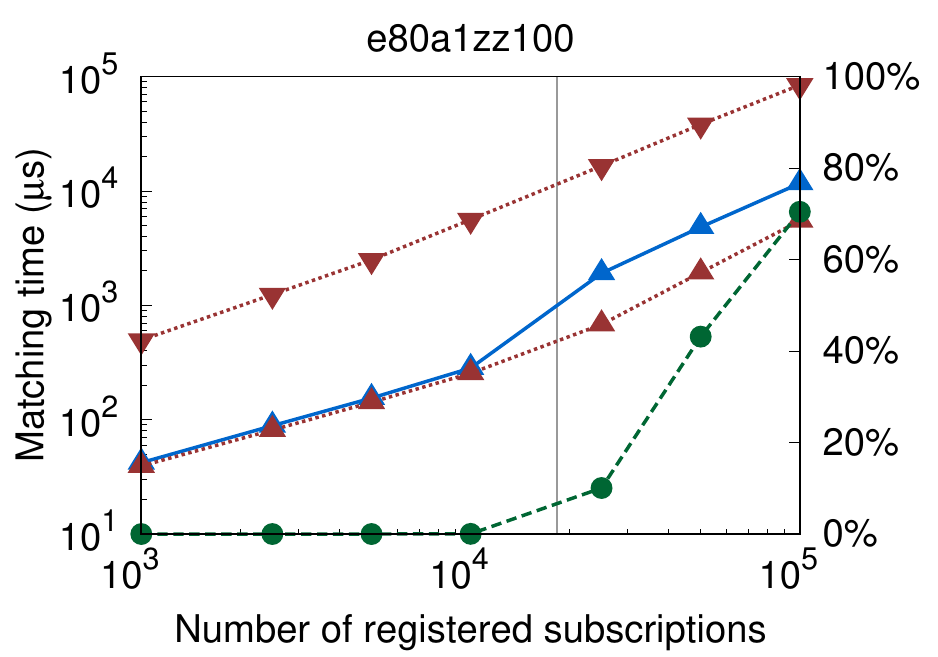}
&
\includegraphics[scale=0.6]{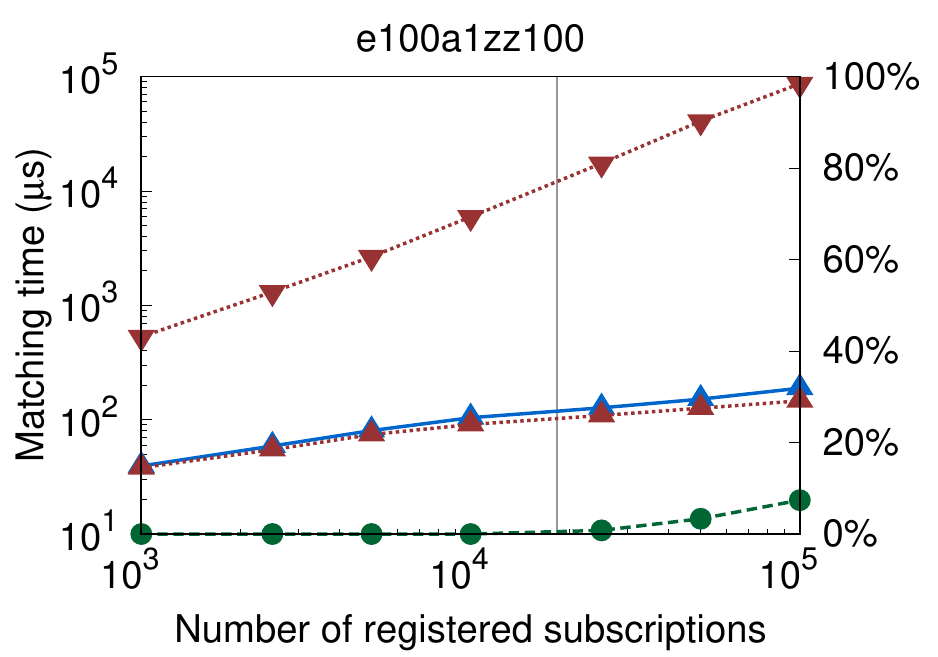}
\end{tabular}
\caption{\label{fig:9graphs}Comparison of different approaches with varying workloads.}
\end{figure*}

Figure~\ref{fig:9graphs} displays separate measurements for each workload running \SYS inside and outside an enclave (both using AES encryption).
We also measured, for each workload, the performance of our implementation of ASPE \cite{Choi2010, DEBS-PF:12} as a baseline for a software-only alternative that does not use enclaves.
We measured only the matching step, and not the encryption or decryption of ASPE messages.
The presented ASPE performance cost was therefore inherent to its matching algorithm, which
grows faster than any other strategy when increasing the size of the subscription database.
The difference is more substantial for the first and last workloads, although it remains close to at least one order of magnitude in all setups.
These observations indicate that the performance penalties of SGX are largely tolerable when considering software-only alternatives for secure filtering, at least when the amount of memory used by the routing engine remains below the EPC assigned size. 
We will further explore this matter below.

Another interesting aspect is the gap between the curves corresponding to execution inside and outside the enclaves.
After approximately 10,000 subscriptions, the versions inside and outside enclaves begin to drift apart due to the number of memory accesses necessary to accomplish every comparison.
At some point, the filtering data does not fit completely in the processor's cache memory and cache misses start to occur more frequently.
When this happens, data must be fetched from system memory and, in the case of enclave executions, it must be decrypted and checked for integrity and freshness.
Moreover, the evicted enclave's cache node must be encrypted before being sent to system memory.
That behavior is consistent with cache miss rates (measured outside the enclaves), which are also reported in Figure~\ref{fig:9graphs}.
We only measured cache misses outside the enclaves, because our Linux version failed to properly monitor the cache performance counters inside enclaves.
Since the code running inside and outside the enclaves is the same, it is reasonable to assume that cache miss rates would be similar.

We finally wanted to observe the performance penalties when exceeding the maximum protected memory size and paging begins to happen.
Since EPC memory is limited, whenever it is full and more space is required, pages must be evicted from the protected area to the main (untrusted) memory.
Accordingly, a page swap occurs every time a previously evicted page is accessed.
Besides the fact that system memory is slower than the processor's cache, which already imposes performance costs, memory page swaps are serviced by the operating system and hence incur an even higher overhead.

\begin{figure}[b!]
\centering
\includegraphics[width=0.48\textwidth]{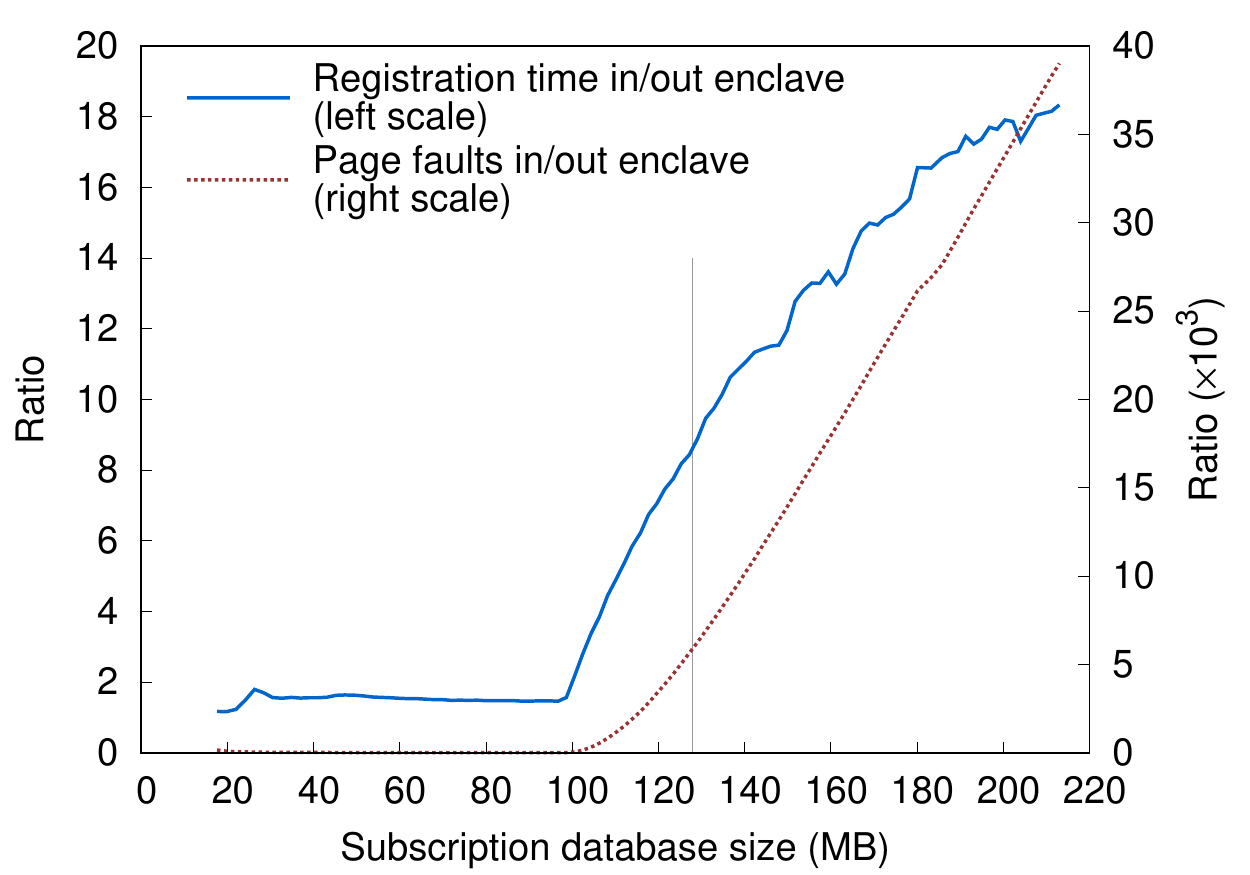}
\caption{\label{fig:blowmemory}Loss in performance when exceeding EPC memory limit.}
~
\end{figure}

Figure~\ref{fig:blowmemory} shows the combined results of two executions when populating the in-memory subscription storage.
In one execution we registered subscriptions inside an enclave, and outside in the other.
We used the workload \emph{e80a1} in plaintext format, and we executed the same registration code in both experiments.
Each point of the graph accounts for an average of 5,000 points, from a set of 500,000 subscriptions.
We plotted the page fault rates observed by dividing their numbers from inside and outside enclaves.
The values measured outside are very large for the largest database size, reaching up to 40,000 more page faults.

We also divided the time it took to register one subscription inside the enclave by the time required outside.
We can clearly see the point where paging starts to take place, when memory consumption reaches just over 90\;MB.
The vertical line shows the EPC memory limit, which comprises both the enclaved application memory and SGX internal data structures. 
At the maximum size of our experiment (213\;MB), registering a subscription inside the enclave took 18 times more time than doing it outside.
These results show that the overhead grows outrageously when paging starts to happen, and they make a strong case for further studies on optimising the memory footprint of applications running inside secure SGX enclaves.

%!TEX root = paper.tex
\section{Related Work}
\label{sec:related}

SGX was introduced only very recently and only few practical algorithms using TEEs have been already proposed.
As far as we could reach in the available literature, \SYS is the first attempt to demonstrate the practical benefits of SGX for privacy-preserving CBR.
This section comments on a couple of related efforts that propose to use SGX in a similar context.

If using SGX for CBR is new, secure content-based publish subscribe is not.
There is a vast amount of literature on the subject, including major surveys \cite{PPPS:Survey:16, Uzunov:CompSec:2016}.
We overview a few relevant existing secure CBR systems (or families thereof) and we discuss the conditions under which an enclaved CBR version can provide an important improvement.

\subsection{SGX as a Secure Building Block}

In a recent paper, Kim et al. \cite{Kim:HotNets:2015} describe three case studies that leverage TEEs in the context of networking applications.
One of them illustrated how SGX can improve on privacy by running parts of software-defined network controllers.
Another case study shows that SGX could be used for protecting Tor onion routers from attacks made by participating nodes.
The last one proposes to use SGX for executing in-network functions.
All case studies share with \SYS the underlying idea of secure execution inside untrusted nodes.

VC3 \cite{Schuster:Security:2015} is a distributed, secure execution environment extending the Apache Hadoop MapReduce framework.
Each worker node hosts an enclaved loader that runs a key exchange protocol, and decrypts and executes map/reduce functions.
VC3 guarantees global integrity by generating work summaries within enclaved workers that can be user-verified after the end of a job.
The proposed environment cleanly isolates within enclaves the computation made under a well established model.
\SYS shares some design ideas with VC3, but the inherently structured map/reduce communication pattern makes it unpractical for implementing publish/subscribe (while the converse would be possible).

\subsection{Secure Content-based Publish/Subscribe}

Content-based publish/subscribe appears at first sight to be incompatible with privacy preservation, as by definition messages must be filtered based on their content, i.e., the filtering engine should be able to see both the data of publications and subscriptions.
To make the problem even more challenging, many filtering techniques exploit structural properties of the information exchanged between publishers and subscribers.
In particular, by structuring the containment relations between subscriptions, one can build efficient data structures to store subscriptions and match publications.
Containment allows for a significant reduction of the number of subscriptions stored as well as the number of matching evaluations executed per publication.
As a consequence, containment is used in most CBR systems in use today \cite{Carzaniga2001, Chand:SRDS:2004, Li:ICDCS:2005}, which makes them largely incompatible with classical cryptographic techniques for privacy preservation.

A recent survey suggests classifying publish/subscribe systems preserving confidentiality in two categories, depending on whether they leverage traditional security techniques or they rely on special-purpose forms of encryption \cite{PPPS:Survey:16}.

Solutions based on traditional techniques encrypt sensitive information that traverse untrusted domains, hence basically preventing the infrastructure from filtering publications based on their content.
They propose architectures designed to respond to specific security threats, through creative combinations of access control and key management mechanisms.
Most of these solutions integrate elaborated access control models into existing event-based middleware \cite{Bacon:DEBS:2008, Zhao:ICDCS:2006, Jacobsen:USENIX:2007}, organising routing brokers in sets that share keys to encrypt or decrypt data.
Different encryption granularities are used, ranging from single attributes to the entire message.
Key management imposes an important overhead and content-based routing is only possible in routers that have the keys to decrypt the relevant matching attributes.
Although these solutions isolate traffic between mutually distrusting domains, by eventually relying on plaintext matching they discourage their use in untrusted hardware as found in clouds.

In contrast to filtering with plaintext, recent research has also explored the development of specific encryption algorithms for publish/subscribe that allow for a direct match of encrypted publications with encrypted subscriptions.
One such solution, which we evaluated in this paper, is asymmetric scalar product-preserving encryption (ASPE) \cite{Choi2010}, initially introduced for secure query computation on encrypted databases.
Publication attributes and subscriptions constraints are represented as coordinates of multidimensional points.
ASPE is based on an exact relation preserving isomorphism and supports subscription containment, albeit it is vulnerable to known-plaintext attacks.
ASPE is asymmetric, hence the decryption key can be shared without compromising the solution.
Given that ASPE's matching complexity is prohibitively high when using a large number of attributes, researchers have proposed to enhance it with a pre-filtering approach \cite{DEBS-PF:12} that expresses equality constraints using Bloom filters.
This allows for quickly identifying subscriptions that are known not to match the publication as their equality constraint(s) cannot be satisfied.

Li \emph{et al.} \cite{Li:Georgia:2004} proposed an approach supporting interval matching by transforming it into prefix-matching and using a prefix-preserving encryption algorithm that supports containment.
The scheme has limited resistance under attack and requires a shared key between publisher and the subscriber.
Ion \cite{Ion:Securecomm:2010} subsequently devised a similar prefix-matching algorithm derived from attribute-based encryption that presents stronger privacy guarantees but with much higher encryption cost.
Encryption is based on El Gamaal \cite{ElGamal:1985:PKC}, thus eliminating the need for shared keys between participants (although common key material must be retrieved from a trusted authority).

Raiciu \emph{et al.} \cite{Raiciu:Securecomm:2006} proposed a hybrid matching mechanism that uses different encryption techniques according to the types of values and constraints.
It can handle partial string matching using a specialised bloom filter, range matching using a prefix-preserving encryption as proposed by Li \emph{et al.} (described above), and a fast protocol for matching arbitrary subscription functions based on Yao's garbled circuits \cite{Yao:1986:GES}.
The mechanism supports containment but may yield false negatives.
Furthermore, two identical subscriptions encrypt to identical cyphertexts, which is semantically insecure.

Nabeel \emph{et al.} \cite{Nabeel:SACMAT:2012} proposed a solution that can be used with any constraints on numerical values based on homomorphic encryption.
It allows for containment, but requires common knowledge of the parameters needed for homomorphic encryption between publishers and subscribers and it can only match numerical values.

Tariq \emph{et al.} \cite{Tariq:DEBS:2010} proposed a peer-to-peer architecture based on attribute-based encryption, which supports numerical comparisons and prefix/suffix constraints on strings.
Peers can act as publishers or subscribers and publications are delivered through an overlay structured as a set of containment-based trees.
It depends on a central and trusted authority that provides master public keys for publication encryption, and subscriptions present weak confidentiality.

All encryption schemes presented in the solutions above are either heavier or weaker than standard production-level encryption (e.g., AES).
By using symmetric encryption and plaintext matching under trusted execution, \SYS is able to combine the best of both worlds.
It provides a novel scheme that
(i)~supports confidentiality for events and filters;
(ii)~permits publishers to express further constraints about who can access their events;
(iii)~handles filters that can express very complex constraints on events even if brokers are not able to access any information in clear on both events and filters;
(iv)~allows brokers to use state-of-the art containment-based filtering, and finally
(v)~does not require publishers and subscribers to share any keys.

%!TEX root = paper.tex
\section{Conclusion}
\label{sec:conclusion}

We presented the architecture and evaluation of \SYS, a secure content-based routing engine that takes advantage of the trusted execution environment provided by shielded SGX enclaves, a technology available on recently released off-the-shelf Intel processors.
In doing so, we can leverage state-of-the-art techniques for efficient filtering in plaintext, since the trusted perimeter is limited to the CPU die.
Outside that boundary, private data is always encrypted and protected from tampering and replay attacks, even from operating systems, hypervisors, and administrators with physical access to machines.
As a result, we do not suffer from the prohibitive performance and space overheads of software-based secure approaches, such as homomorphic encryption or dedicated algorithms like ASPE.
On the other hand, the safety of our system is entirely based on SGX and it could be compromised by design flaws, hardware bugs, or even trapdoors.

As part of our extensive experiments, we tested the system with nine different workloads and analysed the influence of cache misses and page faults on the code running within secure enclaves.
While both events introduce some overhead (as compared to insecure matching outside the enclave), performance degrades much more heavily with the latter, which occurs when exceeding the available amount of protected memory.
We also compared the performance of SGX against the software-based ASPE alternative and observed that SGX performs systematically better as long as memory usage is kept below 90\;MB.
Performance deterioration when exhausting memory is unsurprising.
Indeed, any application that exceeds the EPC size limit will undergo heavy overheads and scalability restrictions. 
This limitation can be overcome through horizontal scalability or future hardware evolutions of SGX.

Our encouraging results open the way for further research on smart handling of subscriptions and matching algorithms in order to minimise memory footprint and build an enclave-efficient system. 
As future work, we intend to explore the possibilities of efficiently use the enclave border and the memory hierarchy.
For instance, we observed that cache misses starts to happen when we have roughly 10,000 subscriptions, which represents 4.37\;MB of memory storage and corresponds to slightly more than half of the 8\;MB cache in the CPU.
Also, we have shown that paging plays an important role in reducing the enclave performance.
We will work on optimising our data structures to avoid paging and cache misses, by smartly storing and accessing the containment trees, splitting them into enclaved and external parts, and appropriately fitting them into cache lines.
Finally, there are also opportunities to optimise our underlying implementation in order to reduce the frequency of enclave enters/exits (e.g., with OS calls).
We plan to explore different strategies such as using message batching, implementing message exchanges at the enclave border, or even bringing more of the simple OS operations into the enclave itself.
All in all, we expect that these mechanisms and optimisations will contribute to further decrease the overhead of running inside an enclave.

~\\

\section{Acknowledgments}
The research leading to these results has received funding from the European Commission, Information and Communication Technologies, H2020-ICT-2015 under grant agreement number 690111 (SecureCloud project).
Rafael Pires is also sponsored by CNPq, National Counsel of Technological and Scientific Development, Brazil.

\bibliographystyle{abbrv}

\par\leavevmode
\end{document}